\documentclass[runningheads]{llncs}

\usepackage[T1]{fontenc}
\usepackage{graphicx}
\usepackage{cite}
\usepackage{amsmath,amssymb,amsfonts}
\usepackage{algorithmic}
\usepackage{graphicx}
\usepackage{textcomp}
\usepackage{xcolor}
\usepackage{url}\usepackage{tikz}
\usetikzlibrary{shapes,arrows,positioning}
\usepackage{pgfplots}
\usepackage{tcolorbox}
\tcbuselibrary{most}
\usepackage{booktabs}
\usepackage{hyperref}
\begin{document}
\title{A Digital Twin to Improve Cyber-Social Impact}
%
%
\author{Vita Santa Barletta\inst{1}\orcidID{0000-0002-0163-6786} 
\and
Danilo Caivano\inst{1}\orcidID{0000-0001-5719-7447} 
\and
Gabriel Cellammare\inst{1,2}\orcidID{0009-0001-1570-4487}
\and
Andrea Dimauro\inst{1}\orcidID{0009-0004-2137-9547}
\and
Pasquale Mesagna\inst{3}\orcidID{0009-0005-6479-6364}
\and
Massimiliano Morga\inst{4}\orcidID{0009-0005-9558-6888}}
\authorrunning{V.S. Barletta et al.}
%
\institute{University of Bari Aldo Moro \and
IMT School for Advanced Studies Lucca \and
Sapienza University of Rome \and
SER\&Practices \\
\email{\{vita.barletta,danilo.caivano\}@uniba.it},
\email{gabriel.cellammare@imtlucca.it},
\email{andreadimauro2001@gmail.com},
\email{mesagna.2260332@studenti.uniroma1.it},
\email{m.morga@serandp.com}
}
\maketitle              
\begin{abstract}
The increasing complexity of Industrial Control Systems (ICS) and their integration with IT infrastructures has significantly expanded the cyber threat landscape. Traditional cybersecurity models, primarily based on the Confidentiality, Integrity, and Availability (CIA triad), fail to capture the broader societal and operational impacts of cyberattacks. This paper proposes a novel framework that integrates Digital Twin technology with a Cyber-Social Security model to simulate, analyze, and quantify cyber threats in critical infrastructures. Virtualized ICS environments built with Mininet and MiniCPS are used to model attack scenarios and assess their consequences across technical and social dimensions. At the core of the framework lies the Cyber-Social Vulnerability Scoring System (CSVSS), a quantitative metric that extends CVSS with a Social Impact dimension grounded in the Entity and Place pillars, producing a single, reproducible severity score that reflects real-world consequences. Crucially, the framework does not automate risk assessment but keeps the human analyst at its center: a Large Language Model (LLM) performs a first-pass analysis – mapping vulnerabilities and drafting risk reports – while the analyst, supported by explainable AI feedback, validates the findings and retains decision authority over critical cases. Experimental results demonstrate the framework's ability to model and score cyber-social impact, while the human-in-the-loop design is presented as a conceptual principle to preserve oversight.
\end{abstract}


\section{Introduction}
Industrial Control Systems (ICS) play a crucial role in managing critical infrastructure, including energy, transportation, and smart cities. However, their increasing interconnection with IT systems has introduced significant cybersecurity challenges. ICS environments are characterized by legacy protocols, limited security mechanisms, and high availability requirements \cite{Badawy_2024_ICSenvironments}. 
Protocols such as Modbus lack encryption and authentication, making them particularly vulnerable to cyberattacks \cite{Idrissi_2019_ICSchallenge}. Consequently, securing ICS and Operational Technology (OT) environments is crucial, as successful attacks can disrupt operational continuity, essential services, and human safety.

Traditional cybersecurity assessment relies on the CIA triad (Confidentiality, Integrity, Availability). Vulnerability severity is commonly quantified using the CVSS (Common Vulnerability Scoring System) \cite{Mell2007CVSS, FIRST2023}, which assesses exploitation conditions and technical impacts on CIA properties. However, this approach remains focused on the technical domain and does not account for broader socio-technical consequences, such as cognitive manipulation, information degradation, loss of trust, or cascading effects across interconnected social and physical systems.
To address this gap, the concept of Cyber-Social Security (CSS) \cite{Barletta_CSS_XRSalento} extends cybersecurity analysis beyond technical infrastructures by considering four interconnected domains: \textit{cyber}, \textit{physical}, \textit{informational}, and \textit{cognitive}. These domains capture how cyber threats propagate and affect individuals, organizations, and society.
This paper proposes a framework that integrates Digital Twin technology with a Cyber-Social Security model to address these limitations. The Digital Twin replicates complex cyber-physical networks, enabling researchers to study and assess the effects of potential cyberattacks not only on systems and data but also on individuals, organizations, locations, and broader ecosystems.
The main contributions are: (i) a \emph{Cyber-Social Impact Model (CIAEP)} that extends the CIA triad with the Entity and Place dimensions
; (ii) the \emph{Cyber-Social Vulnerability Scoring System (CSVSS)}, an extension of CVSS v3.1 that incorporates the new dimensions into a unified severity score;
(iii) a \emph{Digital Twin framework} supported by a human-in-the-loop analysis process, where a Large Language Model assists analysts in vulnerability mapping and risk-report generation while preserving human decision authority.
The paper is structured as follows. Section \ref{sec:related_work} reviews Digital Twin approaches for cybersecurity. Section \ref{sec:proposed_approach} presents the proposed framework and CSVSS metric. Section \ref{sec:methodology} describes the Digital Twin architecture, while Section \ref{sec:attacksim} introduces the attack scenarios. Section \ref{sec:results}discusses the experimental results, and Section \ref{sec:conclusions} concludes the paper and discusses directions for future research.

\section{Related Work}
\label{sec:related_work}
Digital Twins (DTs) have been increasingly adopted to improve cybersecurity in Industrial Control Systems (ICS), IoT networks, smart grids, automotive systems, and smart cities. A common approach combines physics-based or data-driven simulations with machine learning techniques for anomaly detection, enabling proactive security monitoring and safe attack experimentation \cite{Badawy_2024_ICSenvironments}.

Several studies demonstrate the effectiveness of DTs for intrusion detection. Varghese et al. \cite{Varghese_DT_2022} developed a DT of an industrial filling plant integrated with a stacked ensemble classifier capable of detecting process-aware attacks in near real time. Similarly, Sayghe \cite{Sayghe_DT_2025} applied a DT to a water-treatment SCADA system, achieving high detection accuracy against false data injection, DoS, and command injection attacks. El-Hajj \cite{El-Hajj_DT_2024} combined Eclipse Ditto and Snort in a DT-based testbed, showing effective detection of flood attacks, although reconnaissance scans remained challenging.
Beyond intrusion detection, DTs are increasingly used for attack simulation and threat intelligence generation. Dietz et al. \cite{Dietz_2022} leveraged ICS digital twins to execute attack scenarios and automatically produce Cyber Threat Intelligence (CTI) reports in STIX2.1 format. In the automotive domain, Venkatachalapathy \cite{venkatachalapathy2024strategic} used DTs to replicate attack scenarios and analyze vulnerabilities in connected vehicle systems.

Complementary efforts focus on DT-based testbeds. GRFICSv2 \cite{formby2018} provides a realistic ICS environment supporting representative attacks such as Modbus ARP spoofing and PLC code injection. However, its fixed architecture limits the modeling of heterogeneous scenarios. Recent research further highlights the role of DTs in smart cities, where high-fidelity virtual models enable cybersecurity experimentation without affecting operational infrastructures \cite{smartcityDT}.

Despite these advances, securing ICS environments remains challenging. Most existing solutions are adapted from traditional IT security and do not fully address the operational constraints and unique characteristics of ICS environments \cite{Idrissi_2019_ICSchallenge}. Consequently, many conventional security mechanisms cannot be directly deployed without affecting performance or availability.

These limitations motivate the development of flexible and programmable DT frameworks – such as the Mininet\footnote{https://github.com/mininet/mininet} and MiniCPS\footnote{https://github.com/scy-phy/minicps} – adopted in this work, capable of modeling heterogeneous cyber-physical infrastructures and supporting the analysis of cyber-social consequences across different operational scenarios.

\section{Proposed Approach}
\label{sec:proposed_approach}
This work proposes a framework that couples Digital Twin (DT) technology with a Cyber-Social Security model to simulate, analyze, and quantify cyber threats in Industrial Control Systems (ICS) and cyber-physical environments. 
\paragraph{\textbf{Cyber-Social Impact Model.}}
The CIA triad addresses the technical properties of a resource but is silent on \emph{who} and \emph{what} are ultimately harmed by exploiting a vulnerability, affecting not only data protection but also physical protection and digital lives. We therefore extend the triad with two dimensions: \textit{Entity (E)}: the people, organizations, and operational roles affected by an attack. A compromise affects not only data but the individuals and institutions that depend on it; \textit{Place (P)}: the physical or digital location of the affected information and processes. The same attack can stay local or escalate to critical infrastructure, cities, regions, or even geopolitical factors.
For example, to illustrate the distinction, consider a ransomware attack that encrypts the HMI of an energy plant. The \emph{Entity} dimension captures \emph{who} is harmed – the control-room operators who lose access, the operating organization, and ultimately the citizens depending on the service – whereas the \emph{Place} dimension captures \emph{how far} the effect propagates – from the local HMI node to the critical energy infrastructure and, potentially, to an entire region. 
\paragraph{\textbf{Evaluation of Cyber-Social Impact in Smart City.}}

\begin{figure}
    \centering
    \includegraphics[width=0.6\linewidth]{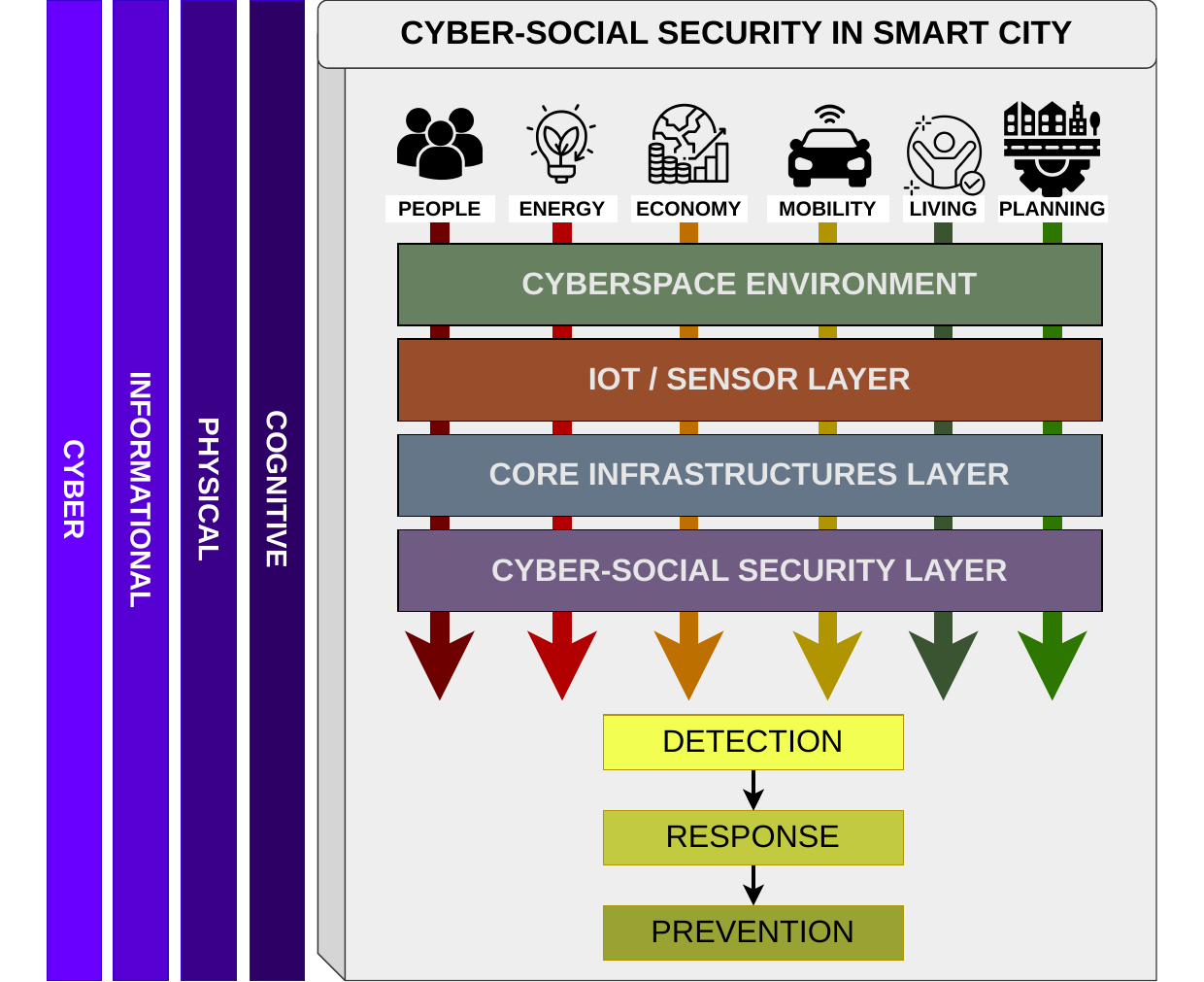}
    \caption{Cyber-Social Security framework for smart cities linking urban domains, technology layers, and cyber-physical-informational-cognitive impacts}
    \label{fig:SmartCityCSS}
\end{figure}

The Cyber-Social Impact Model extends traditional cyber-physical analysis through the Entity and Place dimensions, enabling the assessment of attack effects across interconnected social and spatial scales. To capture the complexity of modern socio-technical environments, the model is instantiated within a Smart City ecosystem \cite{Barletta_SmartProgramManagement}, where digital technologies support urban services, sustainability, and quality of life.
As shown in Fig.~\ref{fig:SmartCityCSS}, the Smart City is described along two axes: a \textit{horizontal dimension}, representing the technological layers (network infrastructure, sensors and devices, management platforms, and applications), and a \textit{vertical dimension}, representing the main urban domains (people, energy, economy, mobility, and livability). A cyberattack originating in one technological layer may propagate across multiple domains, thereby acquiring a cyber-social impact.

The consequences of an attack are further analyzed through four complementary dimensions: \emph{cyber} (systems and services affected), \emph{informational} (information flows and narratives), \emph{physical} (infrastructure and operational processes), and \emph{cognitive} (human perception, trust, and decision-making). Together, these dimensions enable a unified assessment of technical, informational, physical, and human consequences.
However, both CVSS and CIAEP remain fundamentally static models. After each simulation, the affected assets and domains must be identified manually, a process that becomes increasingly complex and time-consuming when analyzing large numbers of attack scenarios within a Digital Twin environment.

\paragraph{\textbf{Cyber-Social Vulnerability Scoring System.}}
\label{subsec:csvss}

\begin{figure*}
    \centering
    \includegraphics[width=0.7\linewidth]{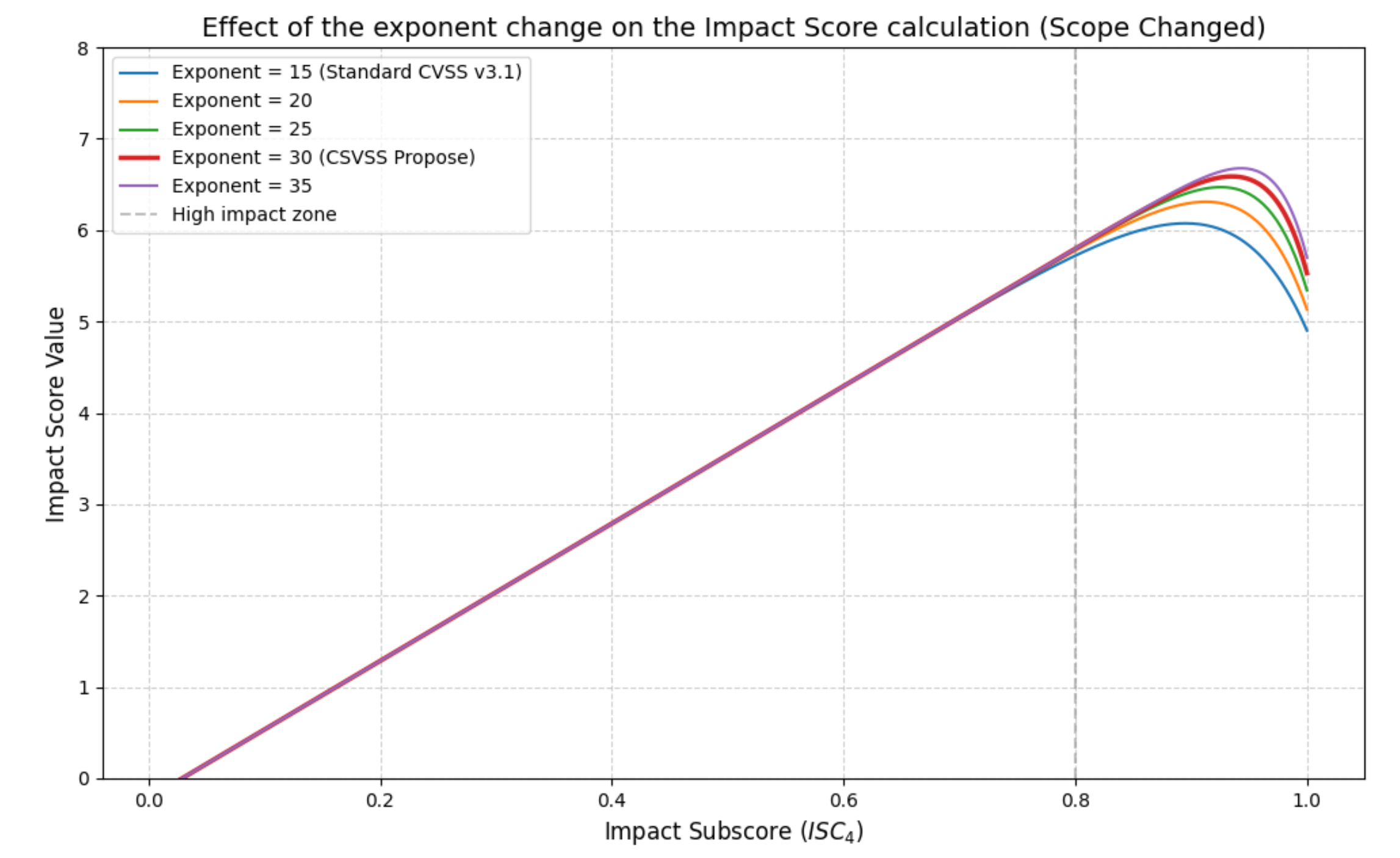}
    \caption{Calibration of the CSVSS corrective term for different exponent values. The comparison highlights how n=30 provides improved discrimination at high-impact levels while preserving monotonicity and maintaining consistency between technical severity and socio-technical consequences.}
    \label{fig:cs-cvss_calibration}
\end{figure*}

CVSS v3.1 derives a Base Score from three impact metrics (C, I, A) combined with exploitability metrics, but it remains confined to the technical sphere.
CSVSS maintains the structure of CVSS v3.1 and adds the Social Impact (SI) dimension as a fourth impact metric, ensuring the final value reflects the social dimension while remaining conceptually and mathematically close to the standard. 
To remove the subjectivity noted in qualitative impact tables, we define explicit anchors for the Entity and Place levels and a deterministic rule that maps them to the scoring system's numeric scale (Table \ref{tab:criteria}). 
Both dimensions are rated on the same three-valued scale adopted by CVSS v3.1 impact metrics: \emph{None} (0), \emph{Low} (0.22)
and \emph{High} (0.56).
 
\begin{table}[t]
\centering
\small
\caption{Assignment criteria for the Entity (E) and Place (P) dimensions.}
\label{tab:criteria}
\scriptsize
\renewcommand{\arraystretch}{1.1}
\begin{tabular}{p{0.10\columnwidth} p{0.37\columnwidth} p{0.37\columnwidth}}
\toprule
Level & Entity (E) & Place (P) \\
\midrule
None (0)  & No human or organizational party affected. & No spatial propagation. \\
Low (0.22)   & A bounded set of individual users; recoverable; no sensitive
       credentials or safety roles involved. &
       Effect confined to a single node or local segment. \\
High (0.56) & Many users/operators, or safety-relevant personnel ,and
       critical organizational roles, sensitive credentials or
       personal data exposed. &
       Propagation to critical infrastructure, multiple smart cities
       domains, or a regional scale. \\
\bottomrule
\end{tabular}
\end{table}
 
The two dimensions are aggregated into a single \emph{Social Impact} (SI) value that feeds the scoring formula. To avoid underestimating social exposure, we adopt a conservative worst-case aggregation:
\begin{equation}
\mathrm{SI} = \max(\mathrm{E},\,\mathrm{P})
\label{eq:si}
\end{equation}
which keeps SI within the admissible CVSS set $\{0, 0.22, 0.56\}$ and
preserves mathematical compatibility with the standard.

All symbols in the equations below follow the CVSS v3.1 Base specification, which builds the score from two metric groups.  
The \emph{Impact} metrics describe the consequence on the affected component and comprise Confidentiality (C), Integrity (I) and Availability (A). 
To these,  CSVSS adds the Social Impact (SI) of Eq. \eqref{eq:si}, and each of the four takes the value None $=0$, Low $=0.22$ or High $=0.56$. The \emph{Scope} (S) is a switch value: it is \emph{Unchanged} when the exploit affects only resources under the security authority of the vulnerable component, and \emph{Changed} when it breaches that authority and reaches a different one. 

The \emph{Exploitability} term, instead, measures how readily the
vulnerability is reached and combines four metrics of the vulnerable component: Attack Vector (AV: Network $0.85$, Adjacent $0.62$, Local $0.55$, Physical $0.20$), Attack Complexity (AC: Low $0.77$, High $0.44$), Privileges Required (PR: None $0.85$, Low $0.62$, High $0.27$ – raised to $0.68$ and $0.50$ when the Scope is Changed) and User Interaction (UI: None $0.85$, Required $0.62$).

The resulting four-pillar Impact Subscore is
\begin{equation}
\label{eq:isc4}
\mathrm{ISC}_4 = 1 - \big[(1-C)(1-I)(1-A)(1-\mathrm{SI})\big]
\end{equation}

As in the standard, the Impact depends on \emph{Scope}:
\begin{align}
\text{Scope Unchanged:}\quad
\mathrm{Impact} &= 6.42 \times \mathrm{ISC}_4
\label{eq:impact_unchanged}
\\
\text{Scope Changed:}\quad
\mathrm{Impact} &=
7.52\,(\mathrm{ISC}_4 - 0.029)
\nonumber
&
\qquad \\&- 3.25\,(\mathrm{ISC}_4 - 0.02)^{30}
\label{eq:impact_changed}
\end{align}
The Exploitability subscore is unchanged from CVSS v3.1:
\begin{equation}
\label{eq:exploitability}
\mathrm{Exploitability} = 8.22 \times \mathrm{AV} \times \mathrm{AC}
        \times \mathrm{PR} \times \mathrm{UI}.
\end{equation}

Impact and Exploitability are combined into the final Base Score following the CVSS v3.1 specification:
\begin{equation}
\mathrm{BaseScore}=
\begin{cases}
0, & \mathrm{Im}\le 0\\[2pt]
\mathrm{Roundup}(\min(x,10)), & S{=}\mathrm{U}\\[2pt]
\mathrm{Roundup}(\min(1.08\,x,10)), & S{=}\mathrm{C}
\end{cases}
\label{eq:basescore}
\end{equation}
where $x=(\mathrm{Impact}+\mathrm{Exploitability})$, while $S{=}\mathrm{U}$ and
$S{=}\mathrm{C}$ denote an unchanged and a changed \emph{Scope}, respectively and $Im$ indicates the \emph{Impact}.
Scope and the cyber-social dimensions capture different aspects of an attack and are therefore assessed independently. Consistent with CVSS v3.1, Scope remains a purely technical attribute describing whether exploitation affects resources beyond the original security authority. In contrast, the Entity and Place dimensions measure the societal reach of the consequences, representing the breadth of affected actors and the spatial or operational extent of the impact.

As a result, these dimensions are not interchangeable. An attack may remain within a single security authority (\textit{Scope: Unchanged}) while affecting a large number of individuals (\textit{Entity: High}), or it may cross multiple security boundaries (\textit{Scope: Changed}) but have limited societal impact (\textit{Entity: Low}, \textit{Place: Low}). Keeping these dimensions independent preserves compatibility with CVSS, avoids double-counting propagation effects, and enables a clearer distinction between technical spread and socio-technical consequences.

The only variation to the original CVSS formulation concerns the exponent used in the Scope-Changed impact function. CVSS v3.1 adopts an exponent of 15; however, because CSVSS aggregates four impact dimensions, high values of $\mathrm{ISC}_4$ occur more frequently. In this case, the corrective subtraction term of Eq.~\eqref{eq:impact_changed} may excessively penalize the score, potentially producing a CSVSS value lower than the corresponding technical CVSS.

Figure \ref{fig:cs-cvss_calibration} illustrates the calibration process. While an exponent of 15 introduces a strong curvature near maximum impact values, reducing score discrimination, an exponent of 30 preserves monotonicity and improves sensitivity in the upper severity range. This adjustment limits the effect of the corrective term and ensures greater consistency between technical severity and the broader socio-technical consequences captured by CSVSS.

Therfore, CSVSS is deliberately built on v3.1 and is \emph{not} a re-implementation of CVSS v4.0 safety. CVSS v4.0 is the first version to account explicitly for Operational Technology and for vulnerabilities with physical \emph{safety} implications through dedicated supplemental and environmental metrics \cite{FIRST2023CVSS40}: the \emph{Social Impact} dimension does not measure direct physical harm to a person or process, but the propagated \emph{societal} consequence of an attack – continuity of essential services, public trust and social stability across smart-city domains. Safety and social impact are complementary lenses; CSVSS contributes the latter, which neither v3.1 nor v4.0 elevates to a first-class component of the score.

\paragraph{\textbf{LLM-based Cyber Risk Analysis.}}

\begin{figure*}
    \centering
    \includegraphics[width=0.8\linewidth]{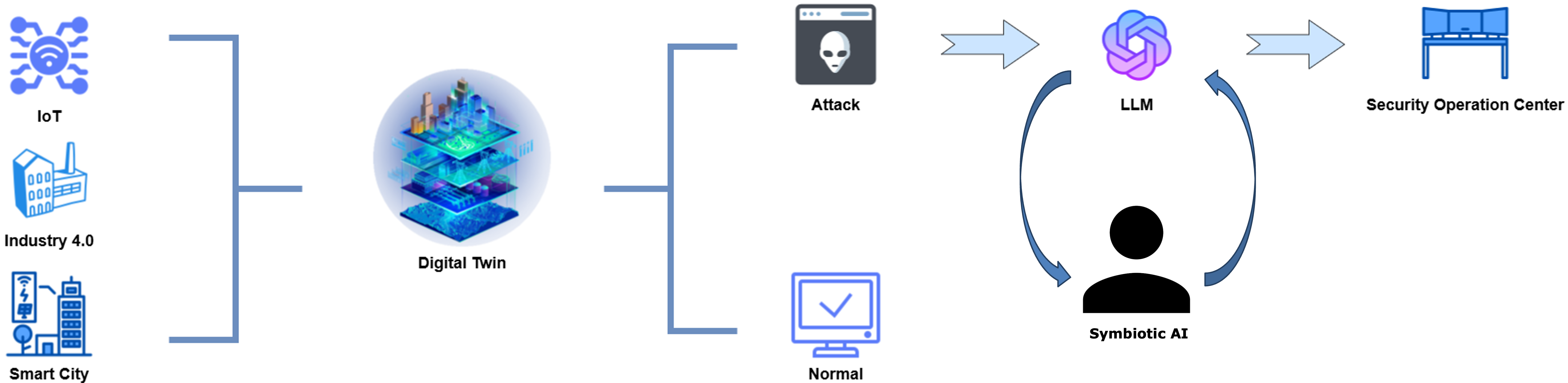}
    \caption{Digital Twin framework for real-world systems: integrating IoT data, attack/normal scenario detection, and LLM-driven symbiotic AI to support Security Operations Center decision-making and risk management.}
    \label{fig:DT_detection}
\end{figure*}

To overcome the static, manual nature of the assessment described above, a Large Language Model (LLM) is integrated into the Digital Twin environment as its core intelligence layer, enabling dynamic, automated analyses (Fig. \ref{fig:DT_detection}).
Unlike traditional Machine Learning approaches, which rely on predefined features and structured datasets, LLMs enable semantic understanding and contextual reasoning over heterogeneous, unstructured data and can autonomously assess the impact of a cyber attack, reducing the time needed to evaluate results.

A general prompt is used across all simulations and then adapted to the specific cyberattack under consideration, possibly including the log files generated within the Digital Twin. The analysis follows a structured pipeline: (i) identification of attack patterns using the CAPEC taxonomy; (ii) mapping of vulnerabilities to CWE categories; (iii) assignment of severity scores using CVSS metrics; and (iv) generation of structured, explainable risk reports.

This pipeline yields automated, context-aware analysis of cyber threats, improving situational awareness and supporting decision-making, and it lets the LLM interpret cyber-physical interactions and their potential cyber-social impact, bridging technical analysis and real-world consequences. The feasibility of leveraging general-purpose generative AI for this task is supported by recent studies \cite{jafarikhah2026descriptionscorellmsquantify}, which confirm that LLMs can automate vulnerability scoring by extracting context directly from textual descriptions, significantly reducing manual overhead. The adopted model is Mistral~7B \cite{jiang2023mistral7b}, a 7-billion-parameter language model: despite its small size, it shows competitive performance among base models with fewer than 13 billion parameters and strong natural-language comprehension, which makes it well suited to automated, contextualized scientific analysis. 

\paragraph{\textbf{Human-in-the-loop and Symbiotic Analysis.}}
\label{subsec:symbiotic}

A design principle of the proposed framework is that the LLM is a symbiotic augmentation layer for the security analyst. This choice reflects the human-centered view that effective human-AI teams exploit the complementary strengths of each partner, letting automation handle routine, well-understood cases while analysts concentrate on tasks that require critical thinking and domain expertise \cite{chhetri2024teaming}. Conceptually, the framework aligns with tiered-autonomy models of human-AI collaboration in security operations \cite{mohsin2025unifiedframeworkhumanai}: rather than a multi-level autonomy taxonomy, it proposes a single human-validation gate, in which the LLM operates with high autonomy on routine, low-criticality findings while critical cases are intended to be escalated to the analyst, who retains decision authority. 

The expected division of tasks is: (i) \emph{AI (implemented)}: synthetic-dataset generation, log triage, mapping of findings to CAPEC/CWE, first-pass CVSS assignment, and drafting of structured reports – the repetitive, high-volume work that drives analyst fatigue; (ii) \emph{Human (by design)}: validation of the AI output, contextual disambiguation, ethical and strategic judgment, and accountability for the final CSVSS verdict in critical scenarios.

This principle is motivated by measured LLM behaviour rather than by convenience: on the task of mapping vulnerabilities to CVSS metrics, recent large-scale evaluation shows that LLMs misclassify a non-negligible and correlated set of cases, and that ensembling yields only marginal gains \cite{jafarikhah2026descriptionscorellmsquantify}. 

\section{Methodology}
\label{sec:methodology}
\begin{figure*}[t]
\centering
\includegraphics[width=1.0\textwidth]{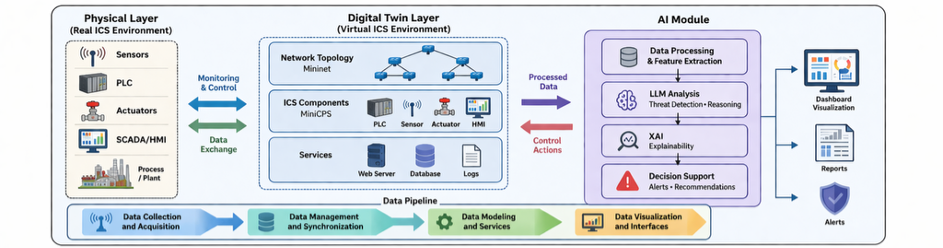}
\caption{Overall architecture of the proposed Digital Twin framework for ICS, integrating physical and digital layers with AI-based analysis.}
\label{fig:dt_arch}
\end{figure*}

\paragraph{\textbf{Digital Twin Architecture.}}
The Digital Twin developed in this work is based on the combined use of Mininet and MiniCPS, enabling the creation of virtualized cyber-physical environments that replicate real-world infrastructures (Fig. \ref{fig:dt_arch}). Mininet provides realistic network emulation, including hosts, switches, and communication links, while MiniCPS extends these capabilities by simulating physical processes and control systems typical of ICS environments \cite{etxezarreta2024use,Somma_2023_ICSframework}.

The architecture is organized into four main layers:
\begin{itemize}
    \item \textit{Data Collection and Acquisition}: This layer gathers data from physical devices, sensors, and simulated industrial components.
    \item \textit{Data Management and Synchronization}: Responsible for handling data flows, ensuring consistency, and synchronizing system states.
    \item \textit{Data Modeling and AI-driven Services}: Enables simulation, data processing, and advanced analysis through Large Language Models (LLMs), supporting semantic interpretation of events and automated report generation. 
    \item \textit{Data Visualization}: Provides interfaces for monitoring system behavior and analyzing results.
\end{itemize}

This layered architecture ensures modularity and adaptability, allowing the DT to be applied across different domains, including IT infrastructures, ICS, and smart city environments \cite{Barletta_SmartProgramManagement}.
The framework was implemented on a moderate workstation (an AMD Ryzen 5700U CPU with 16 GB of DDR4 RAM), demonstrating that the DT does not require high-end hardware. The environment combines VirtualBox, hosting the Ubuntu virtual machine in which the DT runs; Mininet, for network emulation; MiniCPS, extending Mininet with the physical-process and control-logic simulation typical of ICS; and prompt engineering, used to generate the synthetic datasets that feed realistic data flows into the system.

The development of the proposed framework follows a structured methodology composed of two main phases:

\begin{enumerate}
    \item \textit{Digital Twin Design and Implementation}. In this phase, a virtual representation of a cyber-physical infrastructure is designed and implemented. The network topology, communication protocols, and system components are defined using programmable scripts, enabling flexible and reproducible configurations.
    
    \item \textit{LLM-driven Analysis and Enhancement}. Large Language Models are integrated to perform semantic analysis of cyber events, correlate heterogeneous data sources, and generate structured security reports. Unlike traditional Machine Learning approaches, LLMs enable context-aware reasoning, improving the interpretation of complex attack scenarios and supporting the cybersecurity analyst.
\end{enumerate}

This methodology allows for iterative refinement of the DT, progressively increasing the realism and complexity of simulated scenarios. Therefore, the proposed framework emphasizes flexibility, scalability, and realism. By combining network emulation, physical process simulation, and AI-driven analysis, the DT provides a comprehensive environment for cybersecurity experimentation.

\paragraph{\textbf{Network Architecture of the DT.}}

\begin{figure}[t]
\centering
\includegraphics[width=0.8\columnwidth]{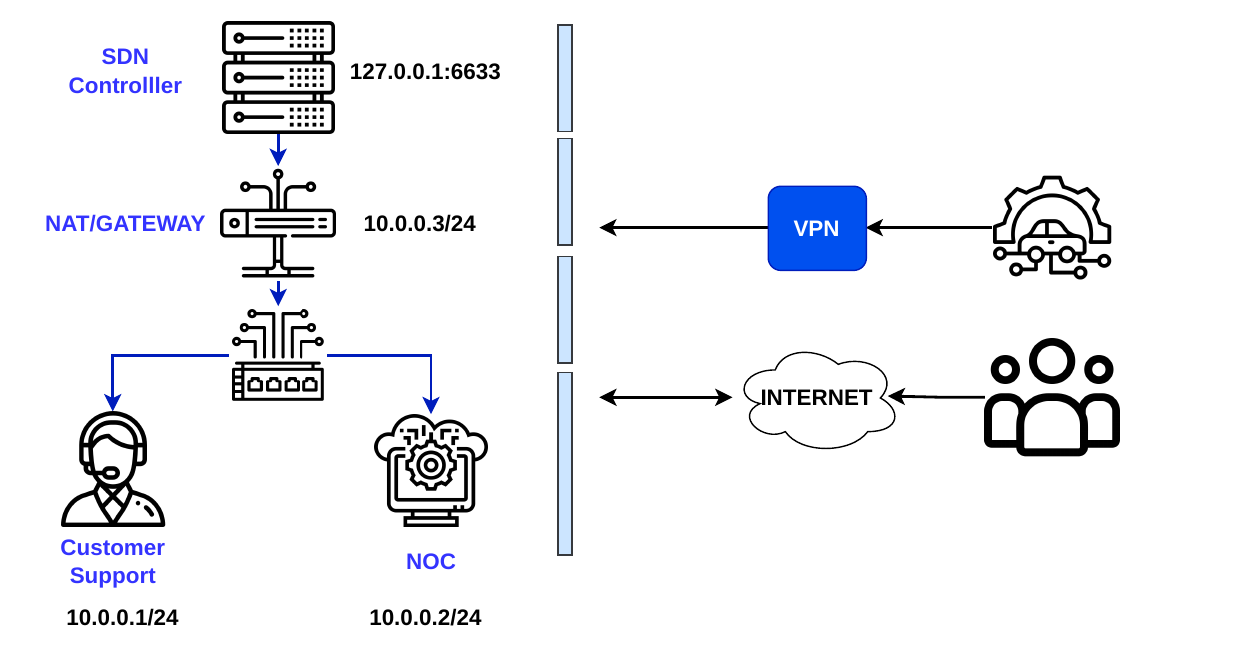}
\caption{Reference network architecture of the industrial Digital Twin (DT). The topology provides a functional abstraction of industrial environment, capturing the main IT and OT components, their interconnections, and the VPN boundary between the two domains. The architecture supports the simulation of cyber-attack propagation from the IT/OT infrastructure toward the smart-city Place dimension.}
\label{fig:networkArchitecture}
\end{figure}

Industrial production is among the sectors most exposed to cyber threats, so the DT is instantiated over a concrete industrial scenario rather than a generic topology. The reference case is inspired by one of the major Italian operators in the energy and manufacturing sectors. The resulting network (Fig.~\ref{fig:networkArchitecture}) is a realistic but fictional industrial scenario that abstracts the key components. Aligned with Smart-Grid conceptual models \cite{kayastha2012smartgrid}, the design realizes the convergence of Information Technology (IT) and Operational Technology (OT) needed to support real-time management and the integration of renewable energy sources. This same IT/OT convergence, however, exposes previously isolated production systems to sophisticated attacks and introduces new critical vulnerabilities \cite{amiri2024itot}. 

The topology is deliberately a \emph{functional abstraction} rather than a faithful replica, for two reasons: first, to bound the simulation complexity, which grows rapidly with the number of nodes; second, to prioritize the generation of immediately analyzable data over detailed reproduction. On the IT side, the network comprises an SDN controller, a NAT/gateway, a switching fabric, a customer support server, and a Network Operations Center, while a VPN boundary separates it from the OT network and external customer access. This layout provides the substrate over which the simulated attacks propagate from the cyber domain into the smart-city Place dimension.

\section{Attack Simulations}
\label{sec:attacksim}
To assess both the resilience of the proposed architecture and the effectiveness of the CSVSS metric, four distinct attack scenarios were implemented within the Digital Twin. The scenarios were deliberately selected to target different layers of the IT/OT infrastructure, so as to observe both the technical degradation and the resulting social impact: (i) \textit{Denial of Service (DoS):} targets the availability of the  Customer support server by flooding the network with a high volume of packets, saturating system resources and allowing the resulting social disruption to be evaluated; (ii) \textit{Cross-Site Scripting (XSS):} reproduces a Broken Access Control condition in the IT network that may lead to unauthorized access within essential service delivery. The choice of XSS is motivated by the high prevalence of this vulnerability class in web security \cite{owasp_top_ten}; (iii) \textit{Ransomware:} models data encryption in order to capture the transition from a localized IT security event to a potential operational halt of essential services \cite{wanasinghe2025dt}; (iv) \textit{Kurtlar.exe malware:} targets VNC management interfaces on ports 5800-5901, exploiting weak credentials to obtain remote control over OT stations and thereby assessing the risk of unauthorized command injection \footnote{\url{https://www.dragos.com/ot-cybersecurity-year-in-review}}.

The outputs of these simulations were subsequently processed using an LLM-based analysis, which automates mapping observed vulnerabilities to the CWE and CAPEC taxonomies and computing the final scores.

\section{Results and Evaluation}
\label{sec:results}

This section reports the results obtained from the four attack scenarios defined in Section \ref{sec:attacksim}, executed within the proposed Digital Twin. The evaluation is organized into four parts: (i) the technical impact analysis, (ii) the cyber-social impact assessment, (iii) the quantitative evaluation through the CSVSS scoring model, and (iv) the LLM-based risk analysis.

\paragraph{\textbf{Technical Impact Analysis.}}

The experiments were carried out in the virtualized ICS environment based on Mininet and MiniCPS. 

The flooding attack produced a marked degradation of service availability, with increased response times and intermittent disruptions in the \textit{Denial of Service} attack. Resource utilization remains consistently high across iterations, confirming the attack's effectiveness in saturating the system's capacity. A representative entry from the server log produced within the DT is:
\begin{tcolorbox}[colback=gray!10, colframe=black!50, sharp corners, boxrule=0.5pt, left=6pt, right=6pt, top=2pt, bottom=2pt, fontupper=\scriptsize\ttfamily]
\begin{itshape}
10.0.0.5 - - [08/May/2025 16:41:38] "GET / HTTP/1.1" 200 -
\end{itshape}
\end{tcolorbox}

The success of the stored \textit{Cross-Site Scripting} chain is documented by three logs. The attacker logs records the payload deployment; the web-server log then shows the injection (\texttt{POST /store\_xss}) followed by two clients reaching the vulnerable endpoint; finally, the stolen-cookies log confirms that the payload executed on the clients, capturing session identifiers and user tokens that could be reused for session hijacking:

\begin{tcolorbox}[
    colback=gray!10,
    colframe=black!50,
    sharp corners,
    boxrule=0.5pt,
    left=6pt,
    right=6pt,
    top=2pt,
    bottom=2pt,
    fontupper=\scriptsize\ttfamily
]
\begin{itshape}
2025-05-16 17:38:26 - attacker - INFO - XSS attack payload deployed \\
10.0.0.3 - - [16/May/2025 17:38:26] "POST /store\_xss HTTP/1.1" 302 \\
10.0.0.5 - - [16/May/2025 17:38:29] "GET /vulnerable\_endpoint HTTP/1.1" 200 \\
10.0.0.5 - - [16/May/2025 17:38:30] "GET /vulnerable\_endpoint HTTP/1.1" 200 \\
Fri May 16 17:35:38 2025 - session\_id=ADMIN\_SESSION\_123; user\_token=ADMIN\_SECRET\_456 \\
Fri May 16 17:38:29 2025 - session\_id=DOWZANW9U8; user\_token=ffxbdhvbq21xen2x
\end{itshape}
\end{tcolorbox}

An initial OSINT phase revealed an SSH service on port 22 as a plausible entry point; the subsequent compromise (through unknown vulnerabilities or social engineering against the node operator) was treated as transparent and not implemented operationally, while the encryption stage was executed and measured. 
The resulting log shows 1002 test files encrypted in 0.16~s, for about 16~KB:
\begin{tcolorbox}[colback=gray!10, colframe=black!50, sharp corners, boxrule=0.5pt, left=6pt, right=6pt, top=2pt, bottom=2pt, fontupper=\scriptsize\ttfamily]
\begin{itshape}
Timestamp, Enc Files, Size (bytes), Time (s) \\
2025-07-06 17:43:37, 1002, 15893, 0.16
\end{itshape}
\end{tcolorbox}
Despite its small scale, the test illustrates how quickly ransomware can act once access is obtained: even a few seconds can cause considerable damage, even in environments equipped with modern defenses.

The simulation demonstrated how the malware Kurtlar.exe can compromise a vulnerable VNC service through brute-force attack, also producing failed attempts against other exposed services. The results of the attack were recorded in the following CSV:
\begin{tcolorbox}[colback=gray!10, colframe=black!50, sharp corners, boxrule=0.5pt, left=6pt, right=6pt, top=2pt, bottom=2pt, fontupper=\scriptsize\ttfamily]
\begin{itshape}
timestamp, target\_name, target\_ip, password, result \\
2025-07-10 10:01:23, host1, 10.0.0.1, 12345, AUTH\_FAILED \\
2025-07-10 10:03:23, host1, 10.0.0.1, 123, AUTH\_SUCCESS \\
2025-07-10 10:02:05, host8, 10.0.0.8, 12345, AUTH\_FAILED \\
2025-07-10 10:03:47, host5, 10.0.0.5, 123, AUTH\_SUCCESS
\end{itshape}
\end{tcolorbox}

Overall, these results confirm that the Digital Twin reproduces the behavior of the four attacks and their technical consequences in a realistic manner.

\paragraph{\textbf{Cyber-Social Impact Assessment}}

The simulated attacks originate in the cyber domain and initially affect the horizontal sensor and network layers of the smart-city architecture, propagating subsequently to vertical domains such as people, the economy, and livability. These cascading effects are interpreted within the Cyber-Social Security paradigm, which integrates both technical and societal perspectives. Accordingly, each scenario is evaluated using the CIAEP model which extends the CIA triad with Entity and Place dimensions.

From a \textit{technical (CIA)} perspective, the DoS attack impacts only availability (C, I = None; A = High). The XSS attack compromises confidentiality through session hijacking and enables integrity violations via impersonation (C, I = High; A = None). The ransomware scenario affects all CIA dimensions through encryption and data exfiltration (C, I, A = High), while the Kurtlar.exe scenario compromises confidentiality, integrity, and availability by enabling full remote control of OT systems (C, I, A = High).

In the \textit{Entity (E)} dimension, the DoS and XSS attacks remain limited to bounded sets of end-users or administrators with fully recoverable effects (E = Low). In contrast, ransomware and Kurtlar.exe impact safety-relevant operators, organizations, and dependent citizens (E = High). Regarding the \textit{Place (P)} dimension, DoS and XSS remain confined to IT services and single web components (P = Low), whereas ransomware and Kurtlar.exe propagate to critical infrastructure and potentially regional-scale operational systems (P = High), with Kurtlar.exe additionally bridging IT, sensor, and OT layers.

The Detection–Response–Prevention lifecycle structures the security response across tactical and strategic levels by integrating these multi-domain effects. Overall, the results highlight how cyberattacks extend beyond technical boundaries to affect real-world entities and environments (Table~\ref{tab:combined}).

\paragraph{\textbf{CSVSS-Based Evaluation}}
While CVSS focuses on the three technical dimensions (Confidentiality, Integrity, Availability), it does not account for the broader societal and physical impact of an attack; CSVSS extends it by introducing the Social Impact dimension built on the Entity and Place pillars, enabling a more comprehensive evaluation.

Table~\ref{tab:combined} reports the two scoring systems across the analyzed scenarios. CSVSS never scores an attack below its CVSS value and raises the severity wherever the social impact is non-negligible, capturing cascading effects that the technical score alone overlooks. For IT-layer attacks, and the Cross-Site Scripting case in particular, the assessment rises when societal reach is taken into account. 

\begin{table}[h]
\caption{Comparison between CVSS and CSVSS scores (left) and attack impact using CIAEP dimensions (right).}
\label{tab:combined}
\scriptsize
\renewcommand{\arraystretch}{1.1}
\centering
\begin{tabular}{lcc}
\toprule
Scenario & CVSS & CSVSS \\
\midrule
DoS         & 7.5 & 8.2 \\
XSS         & 8.9 & 9.8 \\
Ransomware  & 10  & 10  \\
Kurtlar     & 9.4 & 10  \\
\bottomrule
\end{tabular}
\hfill
\begin{tabular}{lccccc}
\toprule
Attack & C & I & A & E & P \\
\midrule
DoS        & None & None & High & Low  & Low  \\
XSS        & High & High & None & Low  & Low  \\
Ransomware & High & High & High & High & High \\
Kurtlar    & High & High & High & High & High \\
\bottomrule
\end{tabular}
\end{table}

Across all scenarios, CSVSS yields a higher severity than CVSS whenever the social impact is non-negligible, in line with the model's objective. The increase is most visible in the Cross-Site Scripting (8.9 to 9.8) and Denial of Service (7.5 to 8.2) scenarios. For Kurtlar, the social dimension pushes the score to the final step (9.4 to 10), whereas for ransomware, the technical severity is already maximal (CVSS = 10), so the social dimension confirms the top-tier rating. The result is \emph{holistic}: the social domain is more than the sum of its parts, which is why the recalibrated, non-linear impact function of Eq.~\eqref{eq:impact_changed} is preferable to a plain additive aggregation.

\paragraph{\textbf{LLM-based Cyber Risk Analysis}}
An LLM is integrated into the system to produce dynamic, automated evaluations: through purpose-built prompts, the model autonomously determines the impact of an attack, drastically reducing the time needed to assess the results. The model adopted is Mistral~7B \cite{jiang2023mistral7b}, a 7-billion-parameter model that achieves state-of-the-art performance among base models with fewer than 13~billion parameters, making it well-suited for automated, context-aware analysis. The following analysis – in the appendix – presents the results of applying LLM-based cyber-social assessment across multiple attack scenarios within the Digital Twin.

\section{Conclusions}
\label{sec:conclusions}

Several limitations bound the present study. 
The Digital Twin is deployed on a modest infrastructure and relies on synthetic data, limiting scalability and realism. Moreover, LLM-based analysis remains sensitive to prompt design and non-deterministic outputs, making human validation essential. The proposed aggregation of Entity and Place dimensions prioritizes transparency and reproducibility, although alternative aggregation methods may be explored in future work. Finally, as Digital Twins introduce additional attack surfaces through interactions between physical and virtual environments, securing these feedback loops remains a critical challenge \cite{mun2024dtsurvey}. Future research will focus on validating the framework in real-world environments, particularly within SOC operations and multi-domain contexts \cite{Barletta_CSS_MDOs}. Further developments include the integration of honeypots, Adversarial Machine Learning (AML), and Explainable AI (XAI) to improve resilience against adversarial attacks and enhance the trustworthiness of automated assessments. The adoption of real-time data streams will also enable predictive simulations and support the evolution of the framework toward SOC-scale deployments.

\section*{Acknowledgment}
This work was partially supported by the following projects: SETH: Security Education and Training Hybrid-Wargame, Avviso “Reti - Sostegno alla ricerca collaborativa” - Codice progetto BM755Y1; CYBER-PREDICT: Cyber vulnerability ranking prediction by prescription, Avviso “Reti - Sostegno alla ricerca collaborativa”, - Codice Progetto DUQVKW0; Patto territoriale "Sistema universitario pugliese" – CUP F61B23000370006;  "AURIGA-EDICO" (Ecological Digital Conversion), under the Regional Regulation n.17/2014 – PIA (art. 26) – Project code: 4WBHY73. The research was conducted in partnership with the Department of Computer Science, University of Bari Aldo Moro (SERLAB – Software Engineering Research Laboratory), focusing on the development of AI/ML-based tools and quality measurement systems for sustainable digital conversion of business processes; "CA-LCAP" (Low-Code Application Platform), under the Programma Operativo FESR 2014-2020 – Contratti di Programma (practice code 4UP72S1). The research was conducted in partnership with the Department of Computer Science, University of Bari, within the framework of a research and development contract (OR3 – Activities A.3.5 and A.3.6).

\bibliographystyle{ieeetr}
\bibliography{references}

\end{document}